\documentclass[namedreferences]{kluwer}
\usepackage{epsf}
\begin{document}
\begin{article}
\begin{opening}
\title{Zodiacal Infrared Foreground Prediction for Space Based Infrared
Interferometer Missions}
\author{M.\surname{Landgraf}\email{Markus.Landgraf@esa.int}}
\author{R.\surname{Jehn}\email{Ruediger.Jehn@esa.int}}
\institute{European Space Agency\\
European Space Operations Center\\
Robert-Bosch-Str.5\\
64293 Darmstadt\\
Germany}
\runningtitle{Zodiacal Infrared Foreground}
\runningauthor{Landgraf and Jehn}

\begin{abstract}
The zodiacal foreground for a highly sensitive space infrared
interferometer is predicted for various observing locations. For the
predictions we use a model that was derived from measurements of
the Cosmic Background Explorer (COBE). We find that at a wavelength of
$10\:{\rm \mu m}$ $96\%$ of the sky is darker than $1\:{MJy}\:{\rm
sr}^{-1}$ for observations in the ecliptic plane at $5\:{\rm AU}$, and
$83\%$ is darker than $0.1\:{\rm MJy}\:{\rm sr}^{-1}$. At $1\:{\rm
AU}$, however, always more than $50\%$ of the sky are brighter than
$1\:{\rm MJy}\:{\rm sr}^{-1}$, even if the observations are made
from $30^\circ$ or $60^\circ$ of latitude above the ecliptic plane, at
$10$ or $20\:{\rm \mu m}$. Thus, according to the employed model, the
foreground reduction by increasing the heliocentric distance of the
observing location is more effective than by increasing the latitude.
\end{abstract}

\keywords{Interplanetary dust, zodiacal light, infrared radiation,
mission analysis}
\end{opening}

\section{Introduction}

The search for extra-solar planets (exo-planets hereafter for brevity)
and possibly for primitive forms of life on them has received much
attention since the mid-1990s and is pursued by NASA and ESA. In its
studies ESA follows a proposal by L\'eger \shortcite{leger96} of a
space infrared interferometer, for which initially five $1\:{\rm
m}$-class telescopes were envisioned. More recent designs of the
interferometer, which was named DARWIN, use six $1.5\:{\rm m}$
telescopes. At NASA a similar instrument with the name of Terrestrial
Planet Finder (TPF) is under development. In addition to the search
for exo-planets such an interferometer could also be used for general
purpose astronomical imaging and spectroscopy. While this paper is
intended to support the mission analysis for DARWIN and TPF, it is
independent from the actual baseline mission scenarios.

One of the main problems for the detection and analysis of Earth-sized
exo-planets is the considerable amount of zodiacal infrared foreground
radiation emitted by the interplanetary dust cloud. In the vicinity of
the Earth either large telescopes have to be used \cite{angel89} or
the observations have to be integrated over long periods of time in
order to sufficiently suppress the zodiacal foreground contribution
\cite{finalreport}. The alternative proposal by L\'eger is to place the
interferometer at a larger heliocentric distance, where the zodiacal
foreground in the $10$-to-$20\:{\rm \mu m}$ range is reduced due to
the lower interplanetary dust density and lower dust
temperatures. There the zodiacal foreground becomes comparable to
other sources of noise, like the radiation leakage from the central
star. As an alternative to larger heliocentric distances it is
conceivable that the zodiacal infrared radiation is reduced at higher
ecliptic latitudes. In order to cover the proposed orbit options, we
investigate heliocentric orbits with aphelia at $1$, $3$, and $5\:{\rm
AU}$ and inclinations of $30^\circ$ and $60^\circ$ \cite{jehn97}.
 
In this paper we determine the zodiacal infrared emission at various
locations (heliocentric distances and ecliptic latitudes) in the solar
system using a model by Kelsall et al.\ \shortcite{kelsall98}. This
model was fit to the data taken by the COBE infrared satellite. We
will discuss our choice of the dust model below.

The result of our analysis is the fraction of the sky where the
zodiacal foreground is of the same order as the other noise
contributors, which is $\dot{N}_{ZL} \approx 2\times 10^4\:{\rm
photons}\:{\rm hour}^{-1}$. At $10\:{\rm \mu m}$ this translates to a
maximum foreground surface brightness of $\approx 0.1\:{\rm
MJy}\:{\rm sr}^{-1}$ for $1.5\:{\rm m}$ telescopes, an interferometric
transmission of $20\:\%$, a spectral resolution of $20$, and a field of
view opening angle of $3.4\:{\rm arcsec}$.

\section{Zodiacal Dust Model}

Up to date, no infrared telescope has been flown beyond $1\:{\rm AU}$
or out of the ecliptic plane. We therefore have to rely on models to
determine the zodiacal infrared foreground at the suggested locations
of the interferometer. If we assume that the dust emits blackbody
radiation, the dust model has to provide the spatial distribution of
the cross section and the temperature of the dust grains in order to
determine the infrared brightness.

There exist a number of interplanetary dust models that are derived
from zodiacal light and infrared observations, as well as in situ
measurements. Our approach here is to use the phenomenological model
by Kelsall et al.\ \shortcite{kelsall98}, that was derived from the
measurements of the COBE/DIRBE instrument. The COBE/DIRBE measurements
represent the most accurate sky survey at infrared wavelengths between
$1.25\:{\rm \mu m}$ and $240\:{\rm \mu m}$ so far. The Kelsall et
al.\ model describes the interplanetary dust cloud as made up of three
components: a smooth lenticular cloud, dust bands generated by
collisions within asteroid families \cite{dermott94a}, and the
Earth-resonant dust ring \cite{reach95}. We only consider the dominant
contribution from the smooth cloud.

The drawback of our approach is that, since COBE was an Earth orbiting
satellite, the model is only well constrained for predictions at
$1\:{\rm AU}$ in the ecliptic. Thus the prediction of an observation
at, for example, $z=1\:{\rm AU}$ above the ecliptic plane along a line
of sight towards the ecliptic north pole is uncertain, because it is
not known how much of the brightness measured by COBE at $z=0$ along
the same line of sight is generated at $0\:{\rm AU} \leq z \leq
1\:{\rm AU}$, and how much at $z>1\:{\rm AU}$. However, the COBE
observations along lines of sight with different ecliptic latitudes do
constrain the vertical profile given by the model, whereas the radial
profile is constrained by lines of sight with different ecliptic
longitudes. The use of the Kelsall et al.\ model has the advantage that
we don't have to make assumptions about the grain size distribution,
because it is based on infrared observations which depend only on the
spatial cross section density, not the spatial particle number
density. The model also does not consider different grain
temperatures for different grain sizes, it applies an {\em average}
grain temperature that depends only on the distance from the Sun.

We summarize the Kelsall et al.\ model. The Infrared brightness
$I_\lambda$ along a line of sight (LOS) is given by
\begin{eqnarray}
I_\lambda & = & \int_{\rm LOS} ds \: n_{\rm cs}(\vec{r}(s)) B_\lambda(T)\\
B_\lambda(T) & = & \frac{2hc^2}{\lambda^5} \frac{1}
{\exp{\frac{hc}{\lambda k T}} - 1}\nonumber\\
T(\vec{r}) & = & T_0 \left|\vec{r}\right|^{-\delta}\nonumber,
\end{eqnarray}
where $h$ is the Planck constant, $c$ the speed of light, and $k$ the
Boltzmann constant. The parameters $T_0=286K$ and $\delta=0.467$ were
fitted to the COBE data. Considering the offset and the tilt of the
zodiacal cloud with respect to the ecliptic plane, the cross section
density $n_{\rm cs}$ is 
\begin{eqnarray}
n_{\rm cs}(\vec{r}) & = & n_0 \left( \frac{r^\prime}{r_0}
\right)^{{-\alpha}} f(\zeta)\\
r^\prime & = & \sqrt{x^{\prime^2} + y^{\prime^2} +
z^{\prime^2}}\nonumber \\
x^\prime & = & x - x_0\nonumber\\
y^\prime & = & y - y_0\nonumber\\
z^\prime & = & z - z_0\nonumber\\
f(\zeta) & = & \exp(-\beta g^\gamma)\nonumber\\
g & = & \left\{
\begin{array}{lll}
\frac{\zeta^2}{2\mu} & \mbox{for} & \zeta < \mu \\
\zeta - \frac{\mu}{2} & \mbox{for} & \zeta \geq \mu 
\end{array}
\right.\\
\zeta & = & \left| \frac{z^{\prime\prime}}{r^\prime}
\right|\nonumber\\
z^{\prime\prime} & = & x^\prime \sin \Omega \sin i - y^\prime \cos
\Omega \sin i + z^\prime \cos i\nonumber
\end{eqnarray}
The parameters are: $r_0 = 1\:{\rm AU}$, $n_0 = 1.13\times
10^{-7}\:{\rm AU}^{-1}$, $\alpha = 1.34$, $\beta=4.14$,
$\gamma=0.942$, $\mu = 0.189$, $i=2.03^\circ$, $\Omega = 77.7^\circ$,
$x_0 = 0.0119\:{\rm AU}$, $y_0 = 0.00548\:{\rm AU}$, and $z_0 =
-0.00215\:{\rm AU}$.

\begin{figure}[ht]
\epsfxsize=\hsize
\epsfbox{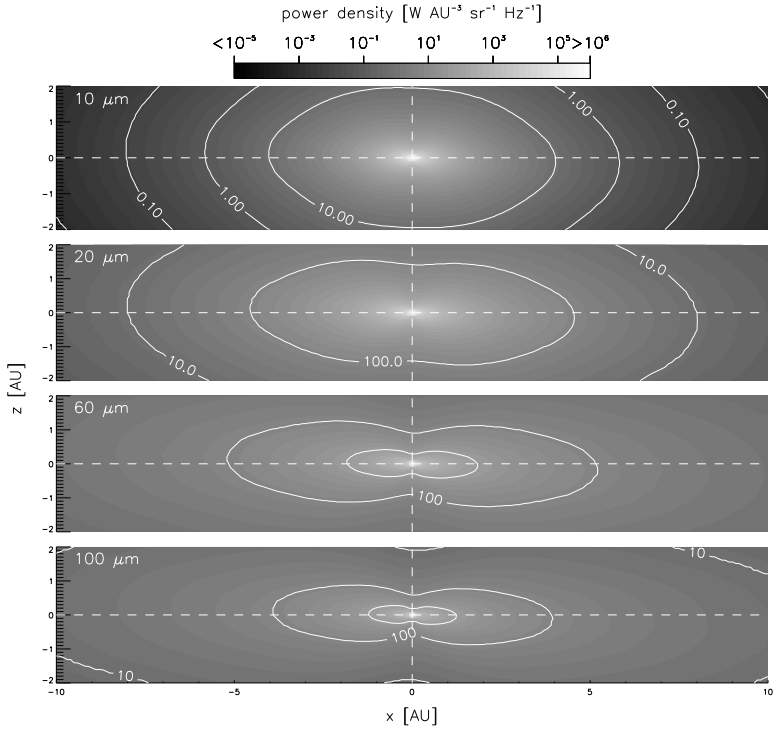}
\caption{\label{fig_cloud} Power distribution radiated by the
interplanetary dust cloud according to the Kelsall et
al.\ \protect\shortcite{kelsall98} model at $10$, $20$, $60$, and
$100\:{\rm \mu m}$. The $z$-direction is perpendicular to the ecliptic
plane and the $x$-axis points towards the vernal equinox.}
\end{figure}

Note that we have not applied a color correction and assume an albedo
of $0$ and an infrared emissivity of $1$. Figure \ref{fig_cloud} shows
the infrared radiation emitted by the smooth interplanetary dust
cloud, predicted by the Kelsall et al.\ model.
At the shorter wavelengths the radiation is emitted from the region
closer to the sun, whereas the more distant regions contribute more to
the radiation at longer wavelengths. Thus, at $10\:{\rm \mu m}$ the
zodiacal foreground is more localized.

\section{Sky Visibility Prediction as Function of the Selected Orbit}

\begin{figure}[ht]
\epsfxsize=\hsize
\epsfbox{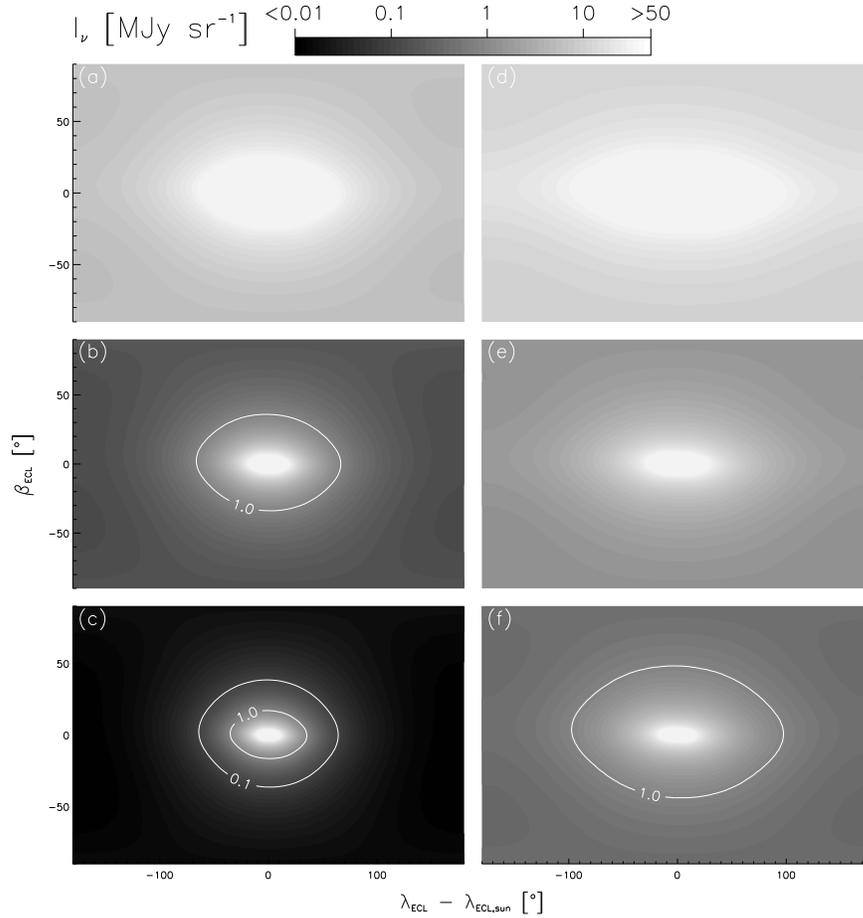}
\caption{\label{fig_sky_radial} Sky maps of the infrared surface
brightness $I_\nu$ of the zodiacal foreground at $10\:{\rm \mu m}$ (a),
(b), (c), and $20\:{\rm \mu m}$ (d), (e), (f). Panels (a) and (d) show
the brightness at an in-ecliptic observing location at $1\:{\rm
AU}$, in (b) and (e) the observation is made at a  heliocentric
distance of $3\:{\rm AU}$, and panels (c) and (f) show the brightness
at $5\:{\rm AU}$. The contour lines show limiting foreground
brightnesses of $0.1$ and $1\:{\rm MJy}\:{\rm sr}^{-1}$.}
\end{figure}

We predict the infrared brightness received from the zodiacal
foreground along lines of sight distributed over the whole sky. To
represent the direction of the line of sight we use two angles : the
solar-relative ecliptic longitude $\lambda_{\rm ECL} - \lambda_{\rm
ECL,\odot}$ and the ecliptic latitude $\beta_{\rm ECL}$. Figure
\ref{fig_sky_radial} shows sky maps of the zodiacal foreground
brightness $I_\nu$ for in-ecliptic orbits at heliocentric distances of
$1$, $3$, and $5\:{\rm AU}$.
Due to the localized nature of the $10\:{\rm \mu m}$ radiation, the
installation of the interferometer at larger heliocentric distances is
very effective in the reduction of the foreground brightness at
$10\:{\rm \mu m}$. The reduction is at least two orders of magnitude in
most regions of the sky. The $20\:{\rm \mu m}$ brightness is less
affected, but still significantly reduced at larger distances.

\begin{figure}[ht]
\epsfxsize=\hsize
\epsfbox{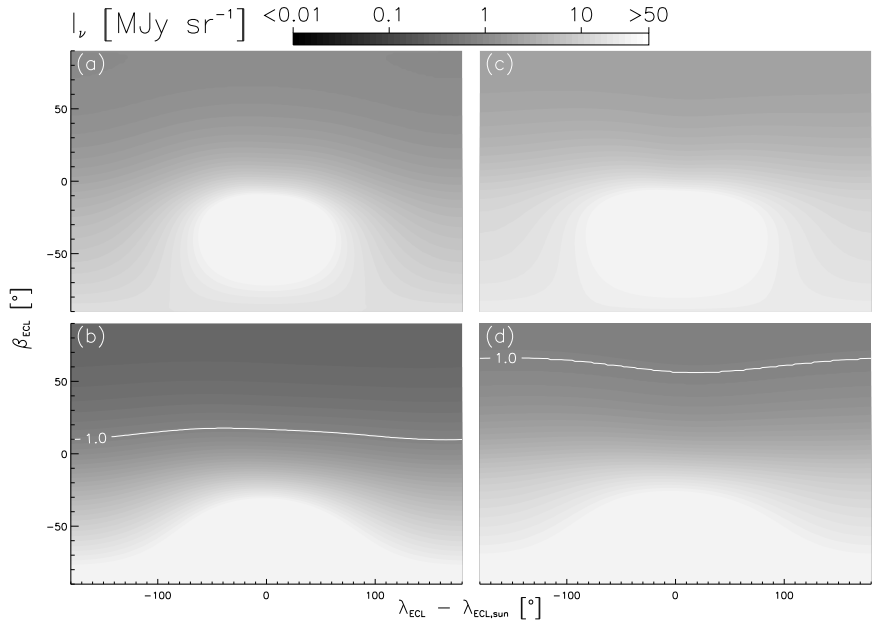}
\caption{\label{fig_sky_incl} Sky maps of the infrared surface
brightness $I_\nu$ of the zodiacal foreground at $10\:{\rm \mu m}$ (a),
(b), and $20\:{\rm \mu m}$ (c), (d). Panels (a) and (c) show the
brightness at an observing location at $30^\circ$ ecliptic latitude
and $1\:{\rm AU}$, and panels (b) and (d) show the brightness at
$60^\circ$ ecliptic latitude. The contour lines show limiting
foreground brightnesses of $0.1$ and $1\:{\rm MJy}\:{\rm sr}^{-1}$.}
\end{figure}

We show sky maps of the foreground at observing locations $1\:{\rm
AU}$ from the Sun, at ecliptic latitudes of $30^\circ$ and
$60^\circ$ in figure \ref{fig_sky_incl}.
According to the Kelsall et al.\ model, the zodiacal infrared
foreground for lines of sight towards the ecliptic north pole is
reduced by one to two orders of magnitude, if an observing location
$60^\circ$ above the ecliptic plane is selected. The reduction factor
is about the same at $10$ and $20\:{\rm \mu m}$. Table
\ref{tab_sky_coverage} shows the fraction of solid angle of the sky
that is darker than $0.1$ or $1\:{\rm MJy}\:{\rm sr}^{-1}$ at $10$
and $20\:{\rm \mu m}$.
\begin{table}[ht]
\caption{\label{tab_sky_coverage} Fraction of the sky with
a zodiacal foreground brightness below $0.1$ and $1\:{\rm
MJy}\:{\rm sr}^{-1}$.}
\begin{tabular*}{\maxfloatwidth}{rrrrrr} 
\hline
& & \multicolumn{2}{c}{fraction of dark sky} &
\multicolumn{2}{c}{fraction of dark sky}\\ 
heliocentric & ecliptic & \multicolumn{2}{c}{($<0.1\:{\rm MJy}\:{\rm
sr}^{-1}$)} &
\multicolumn{2}{c}{($< 1\:{\rm MJy}\:{\rm sr}^{-1}$)}\\
distance$^1$ & latitude$^1$ & $10\:{\rm \mu m}$ &
$20\:{\rm \mu m}$ & $10\:{\rm \mu m}$ &
$20\:{\rm \mu m}$ \\ 
\hline
$1\:{\rm AU}$ & $0^\circ$  & $0\%$  & $0\%$ & $0\%$   & $0\%$  \\
$1\:{\rm AU}$ & $30^\circ$ & $0\%$  & $0\%$ & $0\%$   & $0\%$  \\
$1\:{\rm AU}$ & $60^\circ$ & $0\%$  & $0\%$ & $38\%$  & $6\%$ \\ 
$3\:{\rm AU}$ & $0^\circ$  & $0\%$  & $0\%$ & $83\%$  & $0\%$  \\
$5\:{\rm AU}$ & $0^\circ$  & $83\%$ & $0\%$ & $96\%$  & $70\%$ \\ 
\hline 
\footnotesize{$^1$ of observer's location}
\end{tabular*}
\end{table}

\section{Discussion}

We have predicted the zodiacal foreground brightness at observing
locations in the ecliptic at $1$, $3$, and $5\:{\rm AU}$, as well as
at $30^\circ$ and $60^\circ$ above the ecliptic at $1\:{\rm AU}$ using
the Kelsall et al.\ \shortcite{kelsall98} model. Since the model is
constrained only by observations at $1\:{\rm AU}$ in the ecliptic
plane, the predictions have to be taken with caution, especially at
high latitudes. The validity of the model within the ecliptic plane at
larger heliocentric distances is supported by zodiacal light
observations by imaging photo-polarimeter on board the Pioneer 10
spacecraft
\cite{hanner76}. From the Pioneer measurements it was concluded that
the dust density decreases as $r^{-\nu}$ with heliocentric distance
$r$, and the best-fit value for $\nu$ is $1.0$ to $1.5$. This is in
accord with the radial exponent of $-1.34$ used in the Kelsall et
al.\ model. 

As an alternative to the Kelsall et al.\ model we could have used a
model derived from in situ measurements, like the Divine
\shortcite{divine93} model. The Divine model was fit to, amongst others,
in situ data that was collected at high ecliptic latitudes by the
Ulysses spacecraft. It has to be considered, however, that the in situ
measurements are dominated by particles that are much smaller ($0.1$
to $1\:{\rm \mu m}$) than those that dominantly contribute to the
zodiacal infrared emission ($10$ to $100\:{\rm \mu m}$). Additional
constraints to the high latitude population of interplanetary dust in
the Divine model come from radar meteor measurements, which allow the
determination of meteoroid orbits. Radar meteors give valuable
information about the inclination distribution of interplanetary
meteoroids. However, all observed radar meteors are caused by
Earth-crossing meteoroids, which introduces a bias to the dataset.

In summary it seems worthwhile to do the same prediction, that we have
made with the Kelsall et al.\ model, with the Divine model for
comparison. This would improve our confidence in the projected
zodiacal brightness at the high-latitude observing position. We found,
however, that the Divine model predicts an infrared brightness at
$1\:{\rm AU}$ in the ecliptic that are about a factor of $4$ too low
compared with the COBE measurements along the same line of
sight. Since the Divine model was also fit to observations by the
Infrared Astronomical Satellite (IRAS) \cite{staubach93}, and the
brightness measured by IRAS is in agreement with the COBE
measurements, we conclude that an error in the fit procedure lead to
wrong grain temperatures. Staubach et al.\ \shortcite{staubach93}
report temperatures of $187\:{\rm K}$ for macroscopic ($100\:{\rm \mu
m}$) grains at $1\:{\rm AU}$, which is in deed very low. Only
extremely reflective material would reach such a low equilibrium
temperature at $1\:{\rm AU}$. Kelsall et al.\ \shortcite{kelsall98} use
an average grain temperature at $1\:{\rm AU}$ of $286\:{\rm
K}$. Whatever the solution to the problem is, the discrepancy of the
values shown by Staubach et al.\ \shortcite{staubach93} in their
figures 3 and 4, and the numbers calculated with their equations (1) -
(3) has to be resolved before the Divine model can be used to predict
the infrared brightness.

\section{Conclusion}

Our calculations have shown that an orbit with an aphelion at around
$5\:{\rm AU}$ offers a better sky coverage (i.e.percentage of the sky
for which the zodiacal foreground is below a certain level) than an
inclined orbit at $1\:{\rm AU}$. Even on orbits with inclinations of
$60^\circ$, the infrared emission of the zodiacal light prohibits the
detection of exo-planets in large parts of the sky. In order to reach
such orbital inclinations enormous changes in the orbital velocity are
required that are not feasible with conventional chemical propulsion
systems and that are very costly with electric propulsion engines
\cite{jehn97}. Therefore, for infrared interferometer missions that
are affected by the zodiacal infrared foreground, like the proposed
DARWIN mission, orbits in the ecliptic plane with an aphelion around
$5\:{\rm AU}$ are more promising.

The strong model dependence of our prediction for observing locations
outside $1\:{\rm AU}$ shows, that precursor missions dedicated to make
infrared measurements between $1$ and $5\:{\rm AU}$ are desirable.

Finally it has to be mentioned that during this investigation the
baseline for the DARWIN mission has changed. Now the interferometer is
planned to be installed in one of the quasi-stable Sun-Earth libration
points (probably ${\rm L}_2$), which lie $1.5\times 10^6\:{\rm km}$
from the Earth on the Sun-Earth line, in order to increase the
reliability and to simplify operations and communications
\cite{finalreport}. At this observing location the photon noise from
the zodiacal foreground dominates all other sources of noise like
leaking starlight, detector noise, and galactic background. As can be
seen in figures \ref{fig_sky_radial} (a) and (d) we expect the
zodiacal foreground to be in the order of $10\:{\rm MJy}\:{\rm
sr}^{-1}$ at $10\:{\rm \mu m}$. This means that in order to reduce the
signal to noise ratio to a level where a terrestrial exo-planets can
be detected, the observation of the target star has to be integrated
over a long period of time.

\begin{acknowledgements}
The insightful comments and suggestions of Bill Reach and an anonymous
referee are gratefully acknowledged.
\end{acknowledgements}

\bibliography{dust,mas,darwin}
\bibliographystyle{klunamed}

\end{article}
\end{document}